\documentclass[debug, journal, twocolumn]{rmaa}
\usepackage{natbib}

\usepackage{./aas_macros}

\usepackage{rotating}

\usepackage{paralist}

\usepackage{psfrag,color}

\newcommand{\grsim}{\mathrel{\hbox{\rlap{\hbox{\lower4pt\hbox{$\sim$}}}\hbox{$>$}}}}
\newcommand\TA{\tablenotemark{a}}
\newcommand\TB{\tablenotemark{b}}
\newcommand\TC{\tablenotemark{c}}

\newcommand\hiir{\ion{H}{2} region}
\newcommand\hiirs{\ion{H}{2} regions}
  
\title{The primordial helium abundance and the number of neutrino families} 

\author{
  Peimbert, A.,\altaffilmark{1} 
  Peimbert, M.,\altaffilmark{1}
  and Luridiana, V.\altaffilmark{2,3}}

\altaffiltext{1}{Instituto de Astronom\'{\i}a, Universidad Nacional Aut\'onoma de M\'exico, Mexico.}
\altaffiltext{2}{Instituto de Astrof\'{\i}sica de Canarias, Spain.}
\altaffiltext{3}{Departamento de Astrof\'{\i}sica, Universidad de La Laguna, Spain.}

\shortauthor{Peimbert, Peimbert, \& Luridiana}
\shorttitle{Primordial Helium}

\fulladdresses{
\item Valentiana Luridiana: Instituto de Astrof\'{\i}sica de Canarias, V\'{\i}a L\'actea s/n, 
La Laguna, E-38205, Spain (valentina.luridiana@gmail.com).
\item Antonio Peimbert and Manuel Peimbert: Instituto de Astronom\'\i a, Universidad 
Nacional Aut\'onoma de M\'exico, Apdo. Postal 70-264, M\'exico 04510 D.F., Mexico 
(antonio, peimbert@astro.unam.mx).}

\listofauthors{A. Peimbert, M. Peimbert, \& V. Luridiana}
\indexauthor{Peimbert, A.}
\indexauthor{Peimbert, M.}
\indexauthor{Luridiana, V.}

\abstract{Based on observations of \hiirs\ and the new computations of the recombination coefficients of the 
\ion{He}{1} lines by \citet{por13} we obtain a primordial helium abundance 
by mass of $Y_P = 0.2446 \pm 0.0029$. We consider thirteen sources of error 
for the $Y_P$ determination, some of them are mainly due to systematic 
effects, while the rest are mainly due to statistical effects. We compare our results with other determinations of
$Y_P$ present in the literature. Combining our $Y_P$ value with computations 
of primordial nucleosynthesis we find a number of neutrino species {$ N_{\it eff} = 2.90 \pm 0.22$}, and a neutron 
mean life  {$\tau_{\nu} = 872 \pm 14$(s).}}

\resumen{Basados en observaciones de regiones \ion{H}{2} y los nuevos c\'alculos de los coeficientes de 
recombinaci\'on de las l\'ineas
de \ion{He}{1} debidos a \citet{por13} obtenemos una abundancia primordial de helio por unidad de masa dada 
por $Y_P = 0.2446 \pm 0.0029$. Consideramos trece fuentes de error en la determinaci\'on de $Y_P$, algunas de 
ellas se deben principalmente a effectos sistem\'aticos mientras que el resto se deben principalmente a efectos 
estad\'isticos. Comparamos nuestros resultados con otros valores de $Y_P$ determinados por otros grupos.
Combinando nuestro valor de $Y_P$ con c\'alculos de nucleos\'intesis primordial encontramos que el 
n\'umero efectivo de familias de neutrinos, $N_{\it eff}$, es de $2.90 \pm 0.22$ y que la vida media del
neutr\'on, $\tau_{\nu}$ es de  $ 872 \pm 14 (s)$.
}

\addkeyword{early universe}
\addkeyword{galaxies: abundances}
\addkeyword{galaxies: individual ({SBS~0335--052}, {I~Zw~18}, {Haro~29})}
\addkeyword{galaxies: ISM}
\addkeyword{H II regions}
\addkeyword{ISM: abundances}

\begin{document}
\maketitle

\section{Introduction}
\label{sec:intro}

The determination of $Y_P$ is important for at least the following reasons: a) it 
is one of the pillars of Big Bang cosmology and an accurate determination of 
$Y_P$ permits to test the Standard Big Bang  Nucleosynthesis (SBBN), b) the 
combination of $Y_P$ and $\Delta Y/\Delta O$ is needed to test models of 
galactic chemical evolution, c) the models of stellar evolution require an 
accurate initial $Y$ value, that is given by $Y_P$ plus the additional $Y$ 
produced by galactic chemical evolution, which can be estimated based on the 
observationally determined $\Delta Y/\Delta O$ ratio, d) the determination of the
$Y$ value in metal poor   \hiirs\   requires a deep knowledge of their physical
conditions, in particular the $Y$ determination depends to a significant degree
on their density and temperature distribution, therefore accurate $Y$ determinations 
combined with the assumption of SBBN provide a constraint on the density and 
structure of  {\hiirs}. The first determination of $Y_P$ based on the increase of 
helium with heavy elements was obtained by \citet{pei74}. Historical reviews on 
the determination of the primordial helium abundance have been presented
 by \citet{pei08},  \citet{pag09},  and \citet{ski10}; a recent review on big bang 
nucleosynthesis has been presented by \citet{cyb16}. 

The latest papers on $Y_P$ direct determinations published by each of the three 
main groups working on this subject are:  \citet{ave15}, \citet{izo14} and \citet{pei07} 
(hereinafter Paper 1). In this paper we update the $Y_P$ determination of Paper I 
taking into account, among other aspects, recent advances in the determination of the 
\ion{He}{1} atomic physical parameters by \citet{por13}. We compare our results
with those of Aver et~al.\@ and Izotov et~al.\@ and point out possible explanations 
for the differences among the three determinations.

Paper I may be the most comprehensive attempt to derive the 
primordial helium abundance to date. It includes: a study of 13 sources of error 
involved in this determination; a discussion on the importance of some errors 
that are usually ignored; and a discussion on how to minimize the combined effect 
of all of them. While the study on the error sources presented in Paper I remains 
very relevant, the quantitative value needs to be updated, mostly because of the 
improvements on the theoretical helium recombination coefficients.

\section{Our $Y$ and $Y_P$ determinations}
\label{sec:determin}

\subsection{Tailor made models}
\label{ssec:tailor}

Careful studies of $Y_P$ indicate that the uncertainties in most determinations 
are dominated by systematic errors rather than statistical errors. Increasing 
the number of objects in the samples used to determine $Y_P$ will, of course, 
decrease the statistical errors, however it will not decrease the systematic ones. 

Some systematic errors can be diminished by a careful selection of the 
objects used for the determination as well as by the use of tailor-made models for 
each object. Normal observational procedures, like reddening correction and underlying 
absorption correction, include systematic errors; this occurs because both, 
the reddening law and the underlying absorption correction for the different helium lines, 
are not perfectly known, and any 
error that affects any helium (or hydrogen) line will affect systematically the 
determinations of each object; such systematic effects can be minimized by 
selecting objects with small reddening corrections and \ion{He}{1} large equivalent widths in emission.
Corrections like the ionization correction factor due to the presense 
of neutral helium, $ICF$(He), or the collisional contribution to $I(\mathrm{H}\beta)$ 
depend on the particular objects included in the sample. Since each object is unique, 
there is no such thing as an average $ICF$(He) or a typical $I(\mathrm{H}\beta)$ 
collisional correction for \hiirs; the final error for these effects will be systematic on 
any sample, hence tailor-made models for each object are required.

For the previous reasons we consider that a better $Y_P$ determination can be obtained 
by studying in depth a few \hiirs, rather than by using larger sets of objects without a
tailor-made model for each of them.

\subsection{The new recombination coefficients of the \ion{He}{1} lines}
\label{ssec:coefficients}

To obtain a precise $Y_P$ value it is necesary to have the most accurate atomic 
physics parameters atainable.
\citet{por13} have computed updated effective recombination coefficients for the 
\ion{He}{1} lines that differ from those  by \citet{por05}. 
The new values were computed to correct small
errors in the implementation of case B calculations; they also include a finer 
grid of calculations, useful for high-precision determinations.
The differences between both sets of coefficients are 
small but significant for the determination of $Y_P$. 

From the new atomic data, we present in Table 1
the physical characteristics of our 5 favorite objects derived following the same procedure used in Paper I.

\begin{table*}[!t]\centering
\small
    \setlength{\tabnotewidth}{0.98\linewidth}
  \tablecols{6}
  \caption{Physical Parameters for the \ion{H}{2} Regions} \label{tab:Phys}
 \begin{tabular}{lccccc}
    \toprule
{} & {NGC 346} & {NGC 2363} & {Haro~29} & {SBS 0335-052\TA} & {I Zw 18}  \\
    \midrule
$EW_{em }$(H$\beta$)                                    & $250\pm10$         & $187\pm10$         & $224\pm10$           & $169\pm10$          & $135\pm10$         \\
$EW_{abs}$(H$\beta$)                                    & $2.0\pm0.5$         & $2.0\pm0.5$         & $2.0\pm0.5$           & $2.0\pm0.5$          & $2.9\pm0.5$         \\
$N$(He$^{++}$)/$N$(H$^+$)$(t^2=0.000)$\TB    & $22\pm~2$        & $75\pm12$           & $104\pm9$            & $275\pm8$           & $82\pm23$     \\
$ICF$(He)                                                        & $1.000\pm0.001$ & $0.993\pm0.001$ & $0.9955\pm0.001$ & $0.991\pm0.001$  & $1.000\pm0.001$ \\[2.0ex]
$n_e(t^2=0.000)$                                             & $44\pm17$           & $262\pm77$         & $42\pm50$             & $282\pm44$          & $85\pm84$           \\
$\tau_{3889}(t^2=0.000)$                                 & $0.01\pm0.02$     & $1.14\pm0.41$     & $1.44\pm0.27$       & $2.78\pm0.32$      & $0.06\pm0.05$     \\
$N$(He$^+$)/$N$(H$^+$)$(t^2=0.000)$\TB    & $8333\pm~44$     & $8460\pm149$     & $8421\pm143$       & $8483\pm115$        & $8259\pm314$     \\
$N$(He)/$N$(H)$(t^2=0.000)$\TB                   & $8355\pm47$     & $8476\pm150$     & $8487\pm145$       & $8755\pm117$      & $8341\pm317$     \\
$N$(O)/$N$(H)$(t^2=0.000)$\TB                     & $12\pm2$            & $9\pm1$                & $7\pm1$              & $2.3\pm0.3$         & $1.7\pm0.2$     \\
$O$$(t^2=0.000)$\TC                                      & $14\pm 2$        & $11\pm 1$               & $9\pm 1$           & $2.7\pm 0.3$            & $2.0\pm0.2$       \\[2.0ex]
$t^2$                                                                 & $0.016\pm0.008$ & $0.086\pm0.014$ & $0.029\pm0.007$   & $0.092\pm0.019$  & $0.097\pm0.030$ \\
$n_e(t^2\ne0.000)$                                          & $80\pm31$           & $468\pm122$         & $83\pm65$             & $348\pm52$          & $143\pm131$           \\
$\tau_{3889}(t^2\ne0.000)$                              & $0.03\pm0.03$     & $0.98\pm0.39$     & $1.22\pm0.28$       & $2.75\pm0.35$      & $0.06\pm0.05$     \\
$N$(He$^+$)/$N$(H$^+$)$(t^2\ne0.000)$\TB  & $8271\pm~60$   & $8223\pm150$     & $8314\pm141$      & $8349\pm153$       & $8088\pm350$     \\
$N$(He)/$N$(H)$(t^2\ne0.000)$\TB                & $8293\pm~62$     & $8240\pm151$     & $8380\pm143$      & $8622\pm154$      & $8170\pm352$     \\
$N$(O)/$N$(H)$(t^2\ne0.000)$\TB                 & $13\pm2$             & $19\pm4$             & $9\pm2$                & $5\pm1$                & $5\pm2$     \\
$O  (t^2\ne0.000)$\TC                                   & $16\pm 4$        & $23\pm 8$       & $11 \pm 2$         & $6 \pm 2$        & $6 \pm 3$       \\  
    \bottomrule
	\tabnotetext{a}{Values for the three brightest positions by \citet{izo99}.}
	\tabnotetext{b}{In units of $10^{-5}$.}
	\tabnotetext{c}{Oxygen abundance by mass, in units of $10^{-4}$.}
   \end{tabular}
\end{table*}

\subsection{Updated $Y$ values}
\label{ssec:Yupdate}

To determine $Y_P$ we have to estimate the amount of helium produced by 
the stars during the evolution of the galaxies in our sample; 
to this end we assume that the helium mass increase to 
oxygen mass increase ratio, $\Delta Y/\Delta Z_O$, is constant.
It is possible to determine this ratio self consistently from the points in our 
sample, as is done by \citet{ave15} and \citet{izo14} for their samples. 
We consider that this procedure for a sample as small as ours increases the error in the 
$\Delta Y/\Delta Z_O$ value, instead we use observations of brighter objects of not as 
low metallicity with high quality observations, as well as chemical evolution 
models for galaxies of low mass and metallicity. From \citet{car08}
and \citet{pei10} we obtain $\Delta Y/\Delta Z=1.75$, $Z_O/Z=0.53$ and 
$\Delta Y/\Delta Z_O=3.3 \pm 0.7$.

In Table 2 we present the $Y$ and $Y_P$ determinations for each object of our sample as well as the 
$Y$ values we determined in Paper I. For each determination we have broken 
down the error into its statistical and systematic components: we first 
present the statistical and then the systematic. By taking a weighted average of these 5 $Y_P$ values we obtain 
the updated $Y_P$ value of the sample. The final statistical error amounts to 0.0019, the final 
systematic error amounts to 0.0021;
adding quadratically both components the total error adds up to  $0.0029$.

\begin{table*}[!t]\centering
\small
    \setlength{\tabnotewidth}{0.84\linewidth}
  \tablecols{4}
  \caption{$Y$ and $Y_P$ values(\lowercase{$t^2 \neq 0.000$})} \label{tab:Yvalues}
 \begin{tabular}{lccc}
    \toprule
{} & {$Y$} & {$Y$} & {$Y_P$} \\
{} & {Paper I} & {This Paper\TA} & {This Paper\TB} \\
    \midrule
NGC~346           & $0.2507\pm 0.0027\pm 0.0015$ & $0.2485\pm 0.0027\pm 0.0015$ & $0.2433\pm 0.0028\pm 0.0019$  \\
NGC~2363         & $0.2518\pm 0.0047\pm 0.0020$ & $0.2467\pm 0.0047\pm 0.0020$ & $0.2395\pm 0.0049\pm 0.0026$  \\
Haro~29             & $0.2535\pm 0.0045\pm 0.0017$ & $0.2506\pm 0.0045\pm 0.0017$ & $0.2470\pm 0.0045\pm 0.0019$  \\
SBS~0335--052 & $0.2533\pm 0.0042\pm 0.0042$ & $0.2561\pm 0.0042\pm 0.0042$ & $0.2541\pm 0.0042\pm 0.0042$  \\
I~Zw~18             & $0.2505\pm 0.0081\pm 0.0033$ & $0.2460\pm 0.0081\pm 0.0033$ & $0.2442\pm 0.0081\pm 0.0033$  \\
Sample               & $0.2517\pm 0.0018\pm 0.0021$ & $0.2490\pm 0.0018\pm 0.0019$ & $0.2446\pm 0.0019\pm 0.0021$  \\
    \bottomrule
\tabnotetext{a}{Corrected $Y$ determinations based on the atomic physics 
values presented by Porter et~al.(2013) see text.}
\tabnotetext{b}{Derived from each object under the assumption that 
$\Delta Y/\Delta O = 3.3 \pm 0.7$ see text.}
   \end{tabular}
\end{table*}

It can be seen from Table 2 that the extrapolation from  $Y$ to  $Y_P$ for the objects in our sample
is small and amounts to  $\Delta Y$ = 0.0044.

Once the \ion{He}{1} recombination coefficients have been recomputed \citep{por13}
without the oversights of the previous ones \citep{por05}, we consider that the new 
determinations produce an uncertainty on $Y_P$ of about 0.0010, the value we 
adopted in Paper I.

A thorough discussion on the systematic and statistical errors adopted in our $Y_P$ determination
is presented in Paper I.

\subsection{The fluorescent contribution to the \ion{H}{1} and \ion{He}{1} lines}
\label{ssec:fluorecence}

Nonionizing stellar continua are a potential source of photons for continuum 
pumping of the hydrogen Lyman transitions, the so called case D \citep{lur09}. 
Since these transitions are optically thick, de-excitation 
occurs through higher series lines, in particular excitation to $n_u \ge 3$ produce 
transitions to $n_l \ge 2$. As a result, the emitted flux in the affected 
lines has a fluorescent contribution in addition to the usual recombination one; 
consequently,  Balmer emissivities are systematically enhanced above case B predictions.
Moreover the \ion{He}{1} lines are also enhanced by fluorescence.  To a first approximation
the effect of case D on the \ion{H}{1} lines is compensated by the effect of case D on
the \ion{He}{1} lines. We leave for a future paper an estimate of the importance of case 
D in the $Y_P$ determination.

\section{Comparison with other $Y_{p}$ determinations}
\label{sec:cmparison}

The three best  {$Y_P$} determinations in the literature are presented in Table 3, 
we will call these determinations $Y_P$(\ion{H}{2}).
The three groups use different approaches.
 \citet{izo14} use  28 objects, \citet{ave15} use 15 objects and we use 5.
 We put the main emphasis in the study of the systematic effects and try to reduce them by 
 means of tailor-made models for each object, while \citet{izo14} put the main emphasis on
 the statistical effects, and  \citet{ave15} use a subset of the best objects studied by Izotov
 et~al.. Case D produces  a systematic effect that has not been considered by any of the three groups.

\begin{table*}[!t]\centering
    \setlength{\tabnotewidth}{0.89\linewidth}
  \tablecols{5}
  \caption{$Y_P$ values and predicted equivalent number of neutrino families, $\Delta N_{\nu}$,
   beyond the SBBN} \label{tab:neutrino}
 \begin{tabular}{lcccccc}
    \toprule
$Y_P$(\ion{H}{2}) & $Y_P$(\ion{H}{2}+CMB) & $\Delta N_{\nu}$(\ion{H}{2}) & $\Delta N_{\nu}$(\ion{H}{2}+CMB) & $Y_P$ source \\
    \midrule
 $0.2446\pm0.0029$ & $0.2449\pm0.0029$   & $-0.16\pm0.22$ & $-0.14\pm0.22$  & this paper  \\
 $0.2449\pm0.0040$ & $0.2455\pm0.0040$   & $-0.14\pm0.30$ & $-0.09\pm0.30$ & Aver et~al.\@  (2015) \\
 $0.2551\pm0.0022$ & $0.2550\pm0.0022$   & $+0.63\pm0.16$ & $+0.62\pm0.16$ & Izotov et~al.\@ (2014) \\
  \bottomrule

   \end{tabular}
\end{table*}

While these three determinations should give the same result, there are substancial 
differences in $Y_P$ between that by \citet{izo14} and those by  \citet{ave15} and us, the differences
amount to about $3\sigma$.

One of the main
reasons for the difference between our $Y_P$ determination and that by  \citet{izo14} is due to our 
use of  considerably larger temperature variations than those used by them. They use the direct
method to derive the temperature given by the 4363/5007 [\ion{O}{3}] intensity ratio, 
and assume there are very small temperature variations within each object and that 
$T$(\ion{He}{1}) varies statistically around $T$(\ion{O}{3}).
Alternatively we consider temperature variations to derive  $Y_P$
defined by the $t^2$ parameter \citep{pei67} .  For our sample
we obtain  $\left<t^2\right> = 0.064$. The average  $t^2$ for 27 well observed galactic and extragalactic  \hiirs\ is 
0.044 and the $t^2$ range goes from 0.019 to 0.120 \citep{pei12}; 
this in turn makes $T_e$(\ion{He}{1}) systematically smaller than $T_e$(\ion{O}{3}).

Our  $Y_P$ result is in very good agreement with that of  \citet{ave15}; while they do not include 
temperature inhomogeneities in their calculations, they use a temperaure derived from  \ion{He}{1} lines, 
which, in the prescence of temperature inhomogeneities, remains similar to the mean temperature. The 
main differences between our determination and that of Aver et al. are that we make a deeper 
study of each object (having a tailor-made model for each object), and we include information from 
chemical evolution models regarding the determination of $\Delta Y/\Delta O$ \citep{car08, pei10}. 
On the other hand  \citet{ave15} and \citet{izo14} make use of $\lambda$ 10830 of \ion{He}{1} that 
permits them to have a good handle on the electron density.

Observations of the CMB anisotropy with the Planck satelite can estimate $Y_P$ in two different ways:
1) by determining the number of free electrons in the very early universe from the high order multipole 
moments, we will call this determination {$Y_P$(CMB)},  or 2) by measuring the barionic mass with 
the low order multipole moments and using the
SBBN to determine the resulting $Y_P$ \citep{pla15}. The first method is rather direct and self consistent producing 
$Y_P($CMB$)=0.252\pm0.014$, with unfortunate large error bars; the second method is much more precise 
and yields  $Y_P = 0.2467 \pm 0.0001$, but is sensitive to inputs that go into the SBBN models, 
it is particularly sensitive to the $N_{\nu}$ and  $\tau_{ n}$  adopted values. The first method is a robust independent 
determination of $Y_P$, that is in agreement with all three \hiir\ determinations, and which we have 
used as an additional constraint to our determinations. The second method has internal errors of the order of  0.0001, but 
external errors at least 100 times larger; instead of using this determination to improve the determination 
of $Y_P$, it can be used to try to constrain the external factors to which it is sensitive; Specifically we can 
use the second method in order to constrain the determinations of $N_{\nu}$ and  $\tau_{ n}$.

\section{Determination of $N_{\nu}$ and  \lowercase{$\tau_{ n}$}}
\label{sec:comp-BBN}

The determination  of $Y_P$ based on BBN depends on several input values, 
like the number of neutrino families {$N_{\nu}$} and the neutron life time $\tau_n$.  
In this section we will take advantage of our determination of $Y_P$ to check on 
the validity of these BBN adopted values. With only one additional restriction ($Y_P$), we 
have to fix one of these two physical quantities to estimate the value of the other.
 
\subsection{Determination of $N_{\nu}$ from $Y_P$ and BBN }
\label{ssec:comp-BBN1}

There is still no good agreement on the value of  $\tau_{ n}$, see for example the discussion in \citet{sal16}. There are
three values of  $\tau_{ n}$ that are relevant: a) five determinations based on the bottle method that yield 
 $\tau_{ n} = 879.6 \pm  0.8$(s) \citep{pig15}, b)  two determinations based on the beam method that yield
 $\tau_{ n} = 888.0 \pm  2.1$(s) \citep{pig15}, and c) the average over the best seven measurements presented 
by the Particle Data Group \citet{oli14} that yield $\tau_{ n} = 880.3 \pm  1.1$(s).

We will adopt  $\tau_{ n} = 880.3 \pm 1.1$(s), the recommended value by the
Particle Data Group \citep{oli14}, and the $Y_P$ values derived from \hiirs\ to determine the number of neutrino 
families and we will compare these numbers with that adopted by SBBN to check on the validity of the adopted  number
of neutrino families.

Based on the production of the Z particle by electron-positron collisions in the laboratory and taking into account
the partial heating of neutrinos produced by electron-positron annihilations during BBN, \citet{man05}  find that
$N_{\it eff}$ = 3.046. Therefore the difference between the number of neutrino families and the SBBN number of
neutrino families is given by $\Delta N_{\nu}$ = $N_{\it eff} $ -3.046.

Discussions on the implications for  $N_{\it eff}$  values different from 3.046 have been presented by
\citet{ste13}  and Nollett \& Steigman (2014, 2015).

From the SBBN $N_{\it eff}$ = 3.046 value and the relation $\Delta N_{\it eff} = 75 \Delta Y$ 
\citep{man11}, it follows that our $Y_P$ determination implies 
that $N_{\it eff} = 2.90 \pm 0.22$ and consequently that $\Delta$$N_{\nu}$ amounts to 
-$0.16 \pm 0.22$ (68\% confidence level, CL), result that is in good agreement with SBBN.

In Table 3 we present the   $\Delta N_{\nu}$ values derived from the three  $Y_P$ determinations; 
we also present the {$Y_P$(\ion{H}{2}+CMB)} values that combine the {$Y_P$(\ion{H}{2})} and 
the {$Y_P$(CMB)} values as well as the  $\Delta N_{\nu}$ derived from such $Y_P$ determinations.
 
\citet{izo14} find $Y_P = 0.2551 \pm 0.0022$ that implies an effective number of neutrino families,
$N_{\it eff} = 3.58 \pm 0.25$ (68\% CL), $\pm 0.40$ (95.4\% CL), and $\pm 0.50$ 
(99\% CL) values. This result implies that a non-standard value of $N_{\it eff}$ is preferred 
at the 99\% CL, suggesting the prescence of a fourth neutrino family with a fractional 
contribution to $N_{\it eff}$ at the time of decoupling.

\subsection{Determination of \lowercase{$\tau_{ n}$}  from $Y_P$ and BBN }
\label{ssec:comp-BBN2}

It is possible from the $Y_P$ values and the SBBN to determine  $\tau_{ n}$. Following  \citet{sal16}, we present
in Table 4 the $\tau_{n}$ values obtained from the  $Y_P$ values derived by \citet{izo14}, \citet{ave15} and
ourselves. Also in Table 4 we present the {$\tau_{ n}$(\ion{H}{2}+CMB)} values that combine the {$\tau_{n}$(\ion{H}{2})}
and the $\tau_{n}$(CMB) values.

\begin{table*}[!t]\centering
    \setlength{\tabnotewidth}{0.83\linewidth}
  \tablecols{5}
  \caption{$Y_P$ values and the neutron mean life, \lowercase{$\tau_{ n}$}}
\label{tab:neutron}
 \begin{tabular}{lccccc}
    \toprule
{$Y_P$(\ion{H}{2})} & {$\tau_{n}$(\ion{H}{2})(s)} & {$\tau_{ n}$(\ion{H}{2}+CMB)(s)} & $Y_P$ source \\
    \midrule
 $0.2446\pm0.0029$ & $870\pm14$ & $872\pm 14$ & this paper  \\
 $0.2449\pm0.0040$ & $872\pm19$ & $875\pm 18$ & Aver et~al.\@  (2015) \\
 $0.2551\pm0.0022$ & $921\pm11$ & $921\pm 11$ & Izotov et~al.\@ (2014) \\
  \bottomrule
\
  \end{tabular}
\end{table*}
  
The  $\tau_{ n}$
results by \citet{ave15} and ourselves are within 1$\sigma$ from the average presented by the Particle 
Data Group \citep{oli14}, and while consistent with both, the bottle and the beam, $\tau_{ n}$ determinations, 
they slightly favor the determination based on the bottle method. On the other hand, the determination of 
\citet{izo14} is more than 3$\sigma$ away from both laboratory determinations.
 
The $\tau_{ n}$
values from the three groups derived from  $Y_P$ are within 1$\sigma$ from the result of the SBBN obtained 
by Planck based on the TT, TE, and EE spectra that amounts to  $\tau_{n}({\rm CMB}) = 907 \pm  69$(s) \citep{pla15}. 
The $\tau_{ n}$ Planck result  is independent of the $Y_P$ values derived from  \hiirs.

\section{Conclusions}
\label{sec:con}

We present new $Y$ values for our five favorite \hiirs,\ see Paper I. From these values we obtain 
that $Y_P = 0.2446 \pm 0.0029,$ the main difference with our Paper I result is due to the use of
updated atomic physics parameters. The new estimated error is similar to that of Paper I because the quality
of the data is the same and we are not modifying our estimates of the uncertainty in the systematic errors.

Our $Y_P$ value is consistent with that of \citet{ave15}, but in disagreement with that of  \citet{izo14},
by more than $3{\sigma}$.

$Y_P$ together with BBN can be used to put constraints on  $N_{\nu}$ and $\tau_{ n}$.

The adoption of $\tau_{ n} = 880.3 \pm 1.1$(s) and our $Y_P$ value imply that $N_{\it eff} = 2.90 \pm 0.22$,
consistent with three neutrino families but not with 4 neutrino families.

The adoption of $N_{\it eff}$ = 3.046 and our  $Y_P$ value imply that $\tau_{ n} = 872 \pm 14$(s),
consistent with both high and low values of  $\tau_{ n}$ in the literature.

An increase on the quality of the $Y_P$ determination from  \hiirs\  will provide stronger constraints on the 
$N_{\nu}$ and  $\tau_{ n}$ values.

We are grateful to Gary Ferland, Ryan Porter, and Gary Steigman for fruitful discussions. 
We are also grateful to the anonymous referee for a careful reading of the manuscript 
and some excelent suggestions. A. P., M. P., and V. L. received partial support from 
CONACyT grant 241732. A. P. and M. P. also recieved partial support from 
DGAPA-PAPIIT grant IN-109716.

\end{document}